\documentclass[lettersize,journal]{IEEEtran}
\usepackage{amsmath,amsfonts}
\usepackage{bm}
\usepackage{algorithmic}
\usepackage{algorithm}
\usepackage{array}
\usepackage[caption=false,font=normalsize,labelfont=sf,textfont=sf]{subfig}
\usepackage{textcomp}
\usepackage{stfloats}
\usepackage{url}
\usepackage{verbatim}
\usepackage{graphicx}
\usepackage{cite}
\usepackage{amsthm}
\usepackage{xcolor}

\newtheorem{remark}{Remark}

\begin{document}

\title{Energy-Efficient Analog Beamforming for RF-WET with Charging Time Constraint}

\author{Osmel Mart\'{i}nez Rosabal,
        Onel L. Alcaraz L\'{o}pez, Hirley Alves
\thanks{Copyright (c) 2015 IEEE. Personal use of this material is permitted. However, permission to use this material for any other purposes must be obtained from the IEEE by sending a request to pubs-permissions@ieee.org}        
\thanks{Osmel Mart\'{i}nez Rosabal,
        Onel L. Alcaraz L\'{o}pez, and Hirley Alves are with the Centre of Wireless Communications (CWC), University of Oulu, Finland. \{osmel.martinezrosabal, onel.alcarazlopez, hirley.alves\}@oulu.fi}
\thanks{This research was supported by the Research Council of Finland (former Academy of Finland) 6G Flagship Programme (Grant Number: 346208) and Grant Number: 348515 and the Finnish Foundation for Technology Promotion.}}

\maketitle

\begin{abstract}
Internet of Things (IoT) sustainability may hinge on radio frequency wireless energy transfer (RF-WET). However, energy-efficient charging strategies are still needed, motivating our work. Specifically, this letter proposes a time division scheme to efficiently charge low-power devices in an IoT network. For this, a multi-antenna power beacon (PB) drives the devices' energy harvesting circuit to the highest power conversion efficiency point via energy beamforming, thus achieving minimum energy consumption. Herein, we adopt the analog multi-antenna architecture due to its low complexity, cost, and energy consumption. The proposal includes a simple yet accurate model for the transfer characteristic of the energy harvesting circuit, enabling the optimization framework. The results evince the effectiveness of our RF-WET strategy over a benchmark scheme where the PB charges all the IoT devices simultaneously. Furthermore, the performance increases with the number of PB antennas.
\end{abstract}

\section{Introduction}\label{sec:intro}
\IEEEPARstart{R}ADIO frequency wireless energy transfer (RF-WET) is key for enabling the next generation of Internet of Things (IoT), including massive IoT, extreme edge devices, and zero-energy devices. As charging does not require physical contact when using RF-WET, one can enhance the devices' mechanical durability, enable waterproof/dustproof designs, and reduce the risk of electric shocks. Moreover, RF-WET enables broadcasting energy to multiple users simultaneously in mobile scenarios, e.g., using unmanned aerial \cite{9698986,9849496} and terrestrial \cite{10081057,9963701} vehicles, and can potentially operate in non-line-of-sight (NLoS) channel conditions. More importantly, RF-WET mitigates the need for battery replacements, reducing network maintenance costs and the risk of environmental pollution \cite{alves2021wireless}. 

The energy consumption of the power beacon (PB) is a key performance indicator in WET-enabled networks. It is desirable minimizing it to reduce operational costs and the network's carbon footprint. Indeed, aiming for global decarbonization, the United Nations' sustainable development goals \cite{UNGA2030Agenda}   call for self-sufficient PBs powered by renewable energy  \cite{9698986,9319211}. However, the availability of ambient energy sources is often erratic and may limit the PBs' energy budget.

Fortunately, the PBs can leverage multi-antenna beamforming to dynamically steer the energy in specific spatial directions and increase the total harvested energy without necessarily requiring higher transmission power levels, thus improving the end-to-end energy conversion efficiency of the RF-WET system. In this regard, analog beamforming provides a more affordable solution regarding hardware complexity, cost, and energy consumption compared to other beamforming implementations \cite{9861037}. This is because, as opposed to its multi-RF chain digital or hybrid counterparts, analog beamforming relies on a single RF chain and analog components for phase shifting the signal. Notice though that this hardware simplicity limits analog beamforming to single-beam transmissions with low spatial resolution at the time.

The most common analog beamforming implementation utilizes a network of phase shifters to adjust the phase of the equal-power antennas' feeding signals. To improve spatial resolution, one can adjust the signal's amplitude at the input of the antennas by using, for instance, variable gain amplifiers at the expense of increased hardware complexity and possibly higher power consumption \cite{Park.2023}. Electronically steerable parasitic array radiators are also appealing for realizing analog beamforming. This architecture consists of an active antenna fed by a single RF chain and multiple parasitic antennas terminated in tunable analog loads \cite{Zhang.2023}. Notice that this approach leverages the strong mutual coupling among antennas to adjust the radiation pattern; therefore, it is not suitable for arrays with widely spaced antennas. Load-modulated arrays overcome this limitation by feeding all the antennas with a centralized power amplifier while controlling the impedance of the coupling network to adjust the transmission pattern. Notice that no RF chain is required and the processing, including the array's current distribution, is carried at baseband \cite{Ataeeshojai.2020}. Unfortunately, the dynamic variations of the loads cause impedance mismatches to the effective array's impedance resulting in unwanted power reflections that can deteriorate the system's performance. Finally, one can resort to switches, instead of phase shifters, to control the set of radiating antennas connected at the power amplifier \cite{Zhang.2018}. Since switches require less control/operation power than phase shifters, this architecture may become more appealing in low-cost/power applications. Despite its hardware simplicity, the switches-based implementation has the poorest spatial resolution compared to the aforementioned implementations.

Meanwhile, notice that the components' non-linearities of the RF-WET chain hinder transmit beamforming optimization for minimum PB's energy consumption. Indeed, the RF energy harvesting (EH) circuits have a non-linear transfer characteristic that is a function of the incident RF power, waveform, and modulation \cite{9502719}. Unfortunately, accounting for all these factors may over-complicate the RF-EH circuit model, which hinders solving the specific WET-related problem. On the other hand, ignoring the non-linearities of RF-EH circuits to facilitate the solution process and providing interesting insights may lead to inaccurate/misleading conclusions.

In this letter, we propose an energy-saving RF-WET strategy in which a multi-antenna PB charges a low-power IoT network using a phase shifters-based analog beamforming. For this purpose, the PB schedules one device at a time, driving the input of the RF-EH circuits to their highest power conversion efficiency point. The main contributions are threefold: i) we propose a novel energy efficiency-oriented approach for modeling the transfer function of the RF-EH circuit in the region of interest, i.e., between the region of maximum conversion efficiency and saturation, which enables our proposed charging optimization procedure; ii) we propose a charging strategy that improves the end-to-end energy conversion efficiency, i.e., the ratio between the total harvested energy by the IoT devices and the energy consumed by the PB; and iii) our results show that, if time division is feasible, it outperforms the strategy of charging all devices simultaneously when they are separated by multiple times the wavelength of the transmit signal. 

\textbf{Notation:} Here, we use boldface lowercase letters to denote column vectors and boldface uppercase letters to denote matrices. $\lVert \mathbf{x} \rVert_p$ denote the $\ell_p$-norm of $\mathbf{x}$, while $(\cdot)^T$, $(\cdot)^H$, $\mathrm{tr}(\cdot)$, and $|\cdot|$ denote the transpose, Hermitian transpose, trace, and absolute value operations, respectively. Also, $\mathcal{O}(\cdot)$ is the big-O notation, which specifies worst-case complexity, whereas $\mathbb{C}$ denotes the set of complex numbers with imaginary unit $j = \sqrt{-1}$. Finally, $\mathbb{E}[\cdot]$ is the expectation operator.

\section{System model and problem formulation}\label{sec:systemANDproblem}
\begin{figure}[t!]
    \centering
    \includegraphics[width = .7\columnwidth]{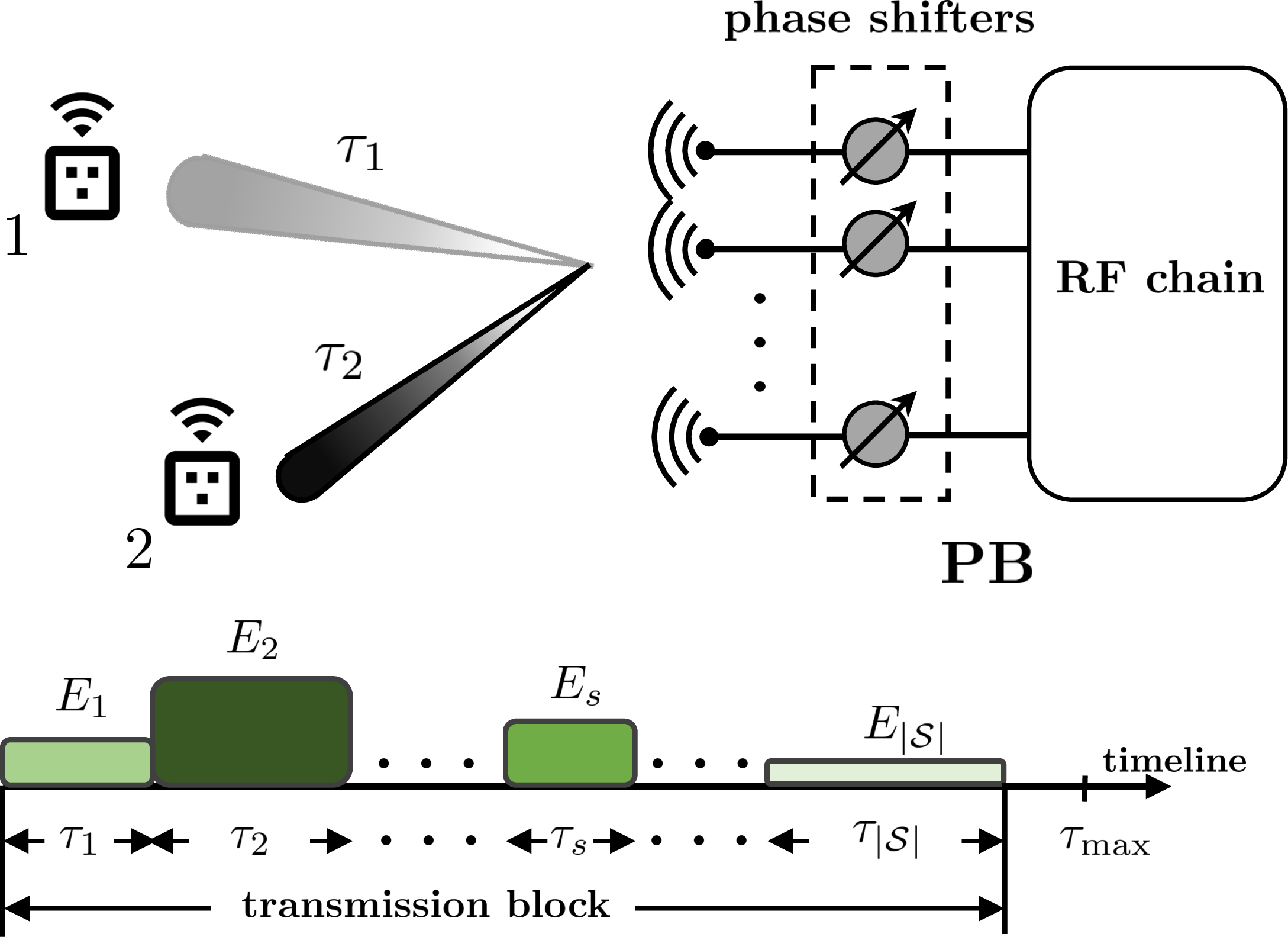}
    \vspace{-3mm}
    \caption{The multi-antenna PB employs a time division strategy to charge the IoT network via analog beamforming.}
    \vspace{-3mm}
    \label{fig:systemModel}
\end{figure}

Consider the scenario where a multi-antenna PB charges a set $\mathcal{S}$ of single-antenna IoT devices. Herein, $\mathbf{h}_s \in \mathbb{C}^{N \times 1}$, where $N$ is the number of transmit antennas, captures the channel coefficients between the PB's antennas and the $s^\mathrm{th}$ device. The system operates over quasi-static fading channels, for which the channel coefficients $\{\mathbf{h}_s\}$, assumed known, remain approximately constant over a transmission block with independent realizations between blocks. Moreover, as Fig.~\ref{fig:systemModel} illustrates, the PB uses a time division charging strategy in which only one device is scheduled, i.e., dedicatedly charged, at a time. Hence, during a timeslot of duration $\tau_s$, the PB transmits a normalized signal $x_s$, i.e., $\mathbb{E}[x_s^H x_s] = 1$, to charge the corresponding $s^\mathrm{th}$ device, whose receive signal is $\mathbf{h}_s^H \mathbf{w}_s x_s$ for a given analog precoder $\mathbf{w}_s \in \mathbb{C}^{N \times 1}$. Thus, by ignoring the received noise power, the harvested energy is 
\begin{equation}\label{eq:harvestedEnergy}
    E_s = \sum_{\forall s^{'} \in \mathcal{S}} \tau_{s'} g \left(  \varpi_{s,s'} \right),
\end{equation}
where $\varpi_{s,s'} \!=\! \mathbb{E}_{x} [|\mathbf{h}^H_s \mathbf{w}_{s'} x_{s'}|^2]
= |\mathbf{h}_s^H \mathbf{w}_{s'}|^2$ is the corresponding incident RF power at the $s^\mathrm{th}$ device when transmitting the signal vector $\mathbf{w}_{s'} x_{s'}$ for charging the device $s'$. Moreover, $g:\mathbb{R}^+ \rightarrow \mathbb{R}^+$ is a generic non-linear transfer function of the RF-EH circuit, which we assume only depends on the incident RF power. Several analytical models are commonly adopted in the literature for $g(\cdot)$, such as piece-wise linear, fractional, cubic fractional, and sigmoidal. We refer the reader to \cite{alves2021wireless} and the references therein for additional information about these models.

Our objective is to minimize the PB's energy consumption while meeting the energy demands $\{E_s^\mathrm{th}\}$ of the IoT devices per transmission block of duration $\tau_\mathrm{max}$. For that, we must find the precoders $\{ \mathbf{w}_s \}$, where $w_{s,n}$ is the $n^\mathrm{th}$ entry of $\mathbf{w}_s$, and per-device charging time $\{\tau_s\}$ such that the required time to charge all devices does not exceed $\tau_\mathrm{max}$. The optimization problem can be formulated as  
\begin{subequations}\label{P1}
    \begin{alignat}{2}
    \mathbf{P1:} &\underset{\{\mathbf{w}_s\},\{\tau_s\}, \{p_s\}}{\mathrm{min.}} \ && \sum_{s \in \mathcal{S}} \tau_s p_s \label{P1a}\\
    &\text{s.t.} \ && E_s \geq E_s^\mathrm{th} , \ \forall s \in \cal S, \label{P1b}\\
    & \ && \sum_{s=1}^{S} \tau_s \leq \tau_\mathrm{max}, \label{P1c} \\
    & \ && |w_{s,n}|^2 \!=\! \frac{p_s}{N}, \ \!\forall s \!\in \!\mathcal{S}, n \!=\! 1,\! \ldots, N, \label{P1d}
    \end{alignat}
\end{subequations}
where $p_s$ is the transmit power associated with the precoder $\mathbf{w}_s$. Notice that \eqref{P1d} corresponds to the constant modulus constraint of the analog precoders. Notably, neither the objective function \eqref{P1a} nor the constraints \eqref{P1b} and \eqref{P1d} are convex\footnote{The Hessian matrix of \eqref{P1a} is neither positive nor negative semidefinite sparse matrix with a single non-zero entry per row/column and with diagonal entries all zero. Besides, $g(\cdot)$ is in general a highly non-linear and non-convex function and thus the constraint \eqref{P1b}. Finally, the constant modulus constraint \eqref{P1d} forms a discrete set and hence is also non-convex.}; therefore, $\mathbf{P1}$ is a non-convex optimization problem.

\vspace{-4mm}
\section{Proposed solution}\label{sec:proposedSolution}
\vspace{-1mm}
Solving $\mathbf{P1}$ requires an analytical expression for the RF-EH circuit's transfer function $g(\cdot)$. In this section, we focus on constructing an accurate mapping that captures the necessary information for the purpose of devising an energy-efficient WET strategy. Without loss of generality, we focus on the point-to-point RF-WET scenario, where the PB charges a single and arbitrary device $s$. Hence, we remove the subindex $s'$ from $\varpi_{s,s'}$ to simplify the notation. Besides, we assume that the transfer characteristics of the EH circuit, as well as relevant key parameters exposed next, have been obtained through direct measurements for a certain operating frequency.

Let $\eta(\varpi_s) = g(\varpi_s)/\varpi_s$ denote the power conversion efficiency function of the rectenna. Besides, let $\varpi_0$ denote the incident RF power that maximizes the function $\eta(\cdot)$. To drive the rectenna at such power level, we consider that the PB utilizes the precoder $\sqrt{p_0/N} \mathbf{u}_s$, where $\mathbf{u}_s$ with $|u_{s,n}| = 1$, $n = 1,\ldots, N$, is the optimal spatial direction for a given $\mathbf{h}_s$. Next, let us consider an arbitrary PB's transmit power $p_s$ for which the resulting incident RF power satisfies $\frac{p_s}{N}|\mathbf{h}_s^H \mathbf{u}_s|^2 \neq \varpi_0$. Finally, let $\tau_0$ and $\tau_s$ denote the required charging times to harvest $E_s$ energy units when the transmit powers are $p_0$ and $p_s$, respectively. Hence, we can state that
\begin{align}
    \tau_0 \frac{p_0}{N} \left| \mathbf{h}_s^H \mathbf{u}_s \right|^2 \ \eta(\varpi_0) &= \tau_s \frac{p_s}{N} \left| \mathbf{h}_s^H \mathbf{u}_s \right|^2 \eta\left(\frac{p_s}{N} \left| \mathbf{h}_s^H \mathbf{u}_s \right|^2\right) \nonumber \\
    \tau_0 p_0  &< \tau_s p_s, \label{eq:optim}
\end{align}
provided that $\eta(\varpi_0)$ is the maximum conversion efficiency. Thus, when $\tau_0$ is feasible, the optimal strategy consists in driving the rectenna to its maximum conversion efficiency point. On the contrary, if $\tau_0$ violates the maximum allowable charging time, the PB must increase its transmit power to deliver $E_s$ energy units within the given deadline. However, exceeding the saturation point $\varpi_\mathrm{sat}$ of the rectenna by increasing the incident RF power requires higher PB's energy consumption and does not provide any additional harvesting power. To show this, let $p_\mathrm{sat}$ denote the transmit power that yields 
$\varpi_\mathrm{sat}$ units of incident power, and let $p_s$ denote any arbitrary transmit power satisfying $\frac{p_s}{N}|\mathbf{h}^H_s \mathbf{u}_s|^2 > \varpi_\mathrm{sat}$. Besides, let $\tau_\mathrm{sat}$ and $\tau_s$ denote the required charging times to harvest $E_s$ energy units when the transmit powers are $p_\mathrm{sat}$ and $p_s$. Hence, we can state 
\begin{align}
    \tau_\mathrm{sat}   \frac{p_\mathrm{sat}}{N} \left| \mathbf{h}_s^H \mathbf{u}_s \right|^2 \eta(\varpi_\mathrm{sat}) &= \tau_s \frac{p_s}{N} \left| \mathbf{h}_s^H \mathbf{u}_s \right|^2 \eta(\frac{p_s}{N} \left| \mathbf{h}_s^H \mathbf{u}_s\right|^2), \nonumber \\
    \tau_\mathrm{sat} p_\mathrm{sat} &< \tau_s p_s, \label{eq:saturation}
\end{align}
since $g(p_s \left| \mathbf{h}_s^H \mathbf{u}_s \right|^2) \approx g(\varpi_\mathrm{sat})$ for $\frac{p_s}{N}|\mathbf{h}^H_s \mathbf{u}_s|^2 > \varpi_\mathrm{sat}$. We summarize this result in the following remark.

\begin{remark}
    For the point-to-point scenario, driving the input of the RF-EH circuit to its highest power conversion efficiency point yields the optimum PB's energy beamforming strategy
\end{remark}

Hence, it is advisable to confine the input of the RF-EH circuit, and consequently the EH function, within the bounds of $[\varpi_0, \varpi_\mathrm{sat}]$ to achieve this goal.

\begin{figure}[t!]
    \centering
    \includegraphics[width=0.9\columnwidth]{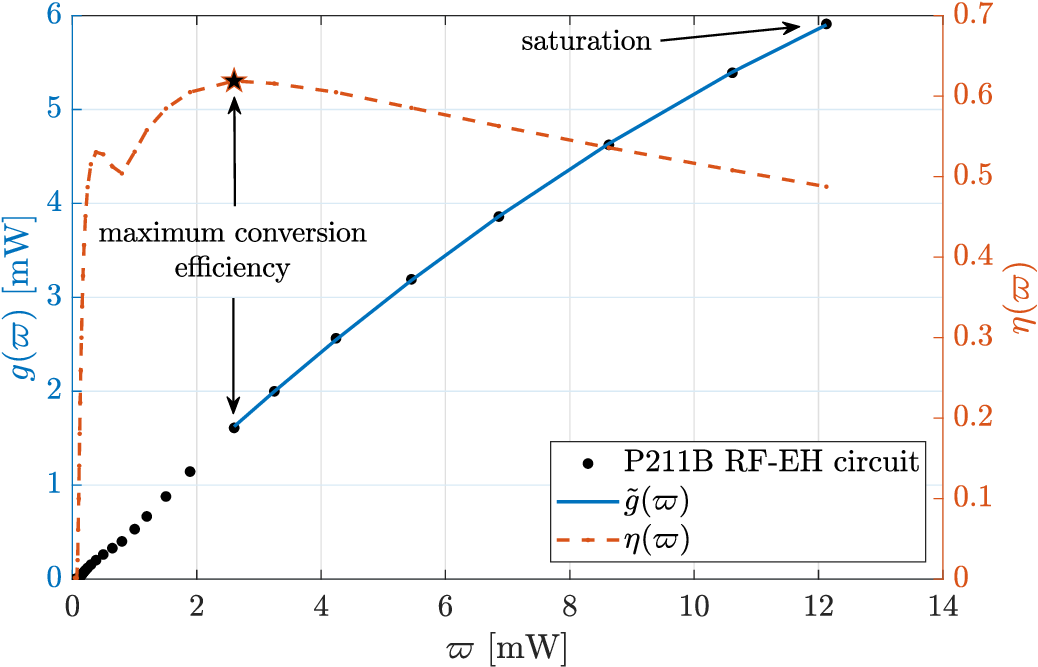}\\
    \vspace{-3mm}
     \caption{Transfer characteristic of the Powercast P2110B RF-EH circuit operating at $868~\text{MHz}$ frequency \cite{powercast_2015}. Herein, $a_2 = -1.952$, $a_1 = 0.663$, and $a_0 = -1.453 \times 10^{-5}$ are obtained by standard curve fitting.}
     \vspace{-4mm}
    \label{fig:EHmodels}
\end{figure}

Herein, we use a second-degree polynomial function, \textit{i.e.}, $\Tilde{g}(\varpi_s) = a_2 \varpi_s^2 + a_1 \varpi_s + a_0$ to fit the transfer function of the Powercast P2110B RF-EH circuit, as Fig.~\ref{fig:EHmodels} depicts. The measurements depicted in Fig.~\ref{fig:EHmodels} were extracted from a power conversion efficiency curve in \cite{powercast_2015} for a maximum input power of around $11~$dBm. Besides, we assume that from that point onward the output of the P2110B RF-EH circuit saturates. Notice that $\Tilde{g}(\cdot)$ accurately models the region of interest and, as we will show in the next subsections, enables our proposed optimization framework. Hereinafter, we utilize $\Tilde{g}(\cdot)$ as our proposed RF-EH model to distinguish from $g(\cdot)$ which models the supported operating range by the RF-EH circuit. In section~\ref{sec:numericalResults}, we explain how the fitting of $g(\cdot)$ is done for performing numerical evaluations.
\vspace{-3mm}
\subsection{Order-agnostic time division}\label{subsec:orderAgnostic}
\vspace{-1mm}
Since the PB establishes a point-to-point connection at each charging timeslot, the optimal per-device analog precoder is $\mathbf{w}_s = \sqrt{\frac{p_s}{N}} \left[
\frac{h_{s,1}}{|h_{s,1}|} \quad \frac{h_{s,2}}{|h_{s,2}|} \quad \cdots \quad \frac{h_{s,N}}{|h_{s,N}|} \right]^T$, which gives an incident RF power of $|\mathbf{h}_s^H \mathbf{w}_s|^2 = \frac{p_s}{N}\lVert \mathbf{h}_s \rVert_1^2$. To ease the mathematical tractability of the problem, we assume that the contribution of the signals $\{ x_{s'} \}, \forall s' \in \mathcal{S}, s' \neq s$, on the $s^\mathrm{th}$ non-scheduled device is negligible, but we consider it in the next subsection as the criterion for establishing a charging order. Therefore, \eqref{eq:harvestedEnergy} reduces to
\begin{align}\label{eq:approxHarvestedEnergy}
    E_s &\approx \tau_s g\left(\left|\mathbf{h}_s^H \mathbf{w}_s \right|^2 \right) = \tau_s g\left(\frac{p_s}{N} \lVert \mathbf{h}_s \rVert^2_1\right).
\end{align}
Then, we can use \eqref{eq:approxHarvestedEnergy} and the quadratic EH function $\Tilde{g}(\cdot)$ to modify the constraint \eqref{P1b} as
\begin{align}
    \tau_s \Tilde{g}\left(\frac{p_s}{N} \lVert \mathbf{h}_s \rVert^2_1\right) &\geq E^\mathrm{th}_s, \nonumber \\ 
    \frac{a_2 p_s^2}{N^2} \lVert \mathbf{h}_s \rVert_1^4 + \frac{a_1 p_s}{N} \lVert \mathbf{h}_s \rVert_1^2 + a_0 &\geq E_s^\mathrm{th} \tau^{-1}_s, \nonumber \\
    c_2 \tau_s^{-1} p_s^{-1} + c_1 p_s + c_0 p_s^{-1} &\leq 1, \label{eq:ehConstraint}
\end{align}
where $c_2 = \frac{E_s^\mathrm{th} N }{a_1 \lVert \mathbf{h}_s \rVert_1^2}$, $c_1 = \frac{a_2 \lVert \mathbf{h}_s \rVert_1^2}{a_1 N}$, and $c_0 = \frac{a_0 N}{a_1 \lVert \mathbf{h} \rVert_1^2}$ are constants. The following result generalizes this procedure for other PB/devices architectures.

\begin{remark}
    Notice that when relying on alternative analog beamforming implementations, e.g., \cite{Park.2023,Zhang.2023,Ataeeshojai.2020,Zhang.2018}, and/or employing multi-antenna RF-EH circuits with RF combining, this procedure still holds and only the precoding/combiner optimization is different.
\end{remark}

Now, we can reformulate $\mathbf{P1}$ as the following geometric programming problem
\begin{subequations}\label{P2}
    \begin{alignat}{3}
    \mathbf{P2:} \ &\underset{\{p_s\}, \{\tau_s\}}{\mathrm{min.}} \ && \sum_{s \in \mathcal{S}} \tau_s p_s \label{P2a}\\
    &\text{s.t.} \ && c_2\tau_s^{-1}p_s^{-1} + c_1p_s + c_0p_s^{-1} \leq 1, \ && \forall s \in \mathcal{S}, \label{P2b} \\
    & \ && w_0 \leq p_s \frac{\lVert \mathbf{h}_s \rVert_1^2}{N} \leq w_\mathrm{sat}, &&\forall s \in \mathcal{S}, \label{P2c} \\
    & \ && \sum_{s = 1}^{S} \tau_s \leq \tau_\mathrm{max},
    \label{P2d}
    \end{alignat}
\end{subequations}

\noindent where the additional constraint \eqref{P2c} guarantees that the incident power remains within the interval $[\varpi_0, \varpi_\mathrm{sat}]$. Although $\mathbf{P2}$ is still non-convex, it can be recast as a convex problem by making the change of variables $\Tilde{\tau}_s = \log{\tau_s}$ and $\Tilde{p_s} = \log{p_s}$ and then apply the function $\log{(\cdot)}$ to the resulting objective and constraints \cite{boyd2004convex}. The resulting optimization problem can be then efficiently solved with accuracy error $\epsilon$ using, e.g., interior-point methods, for which the required number of iterations is in the order of $\mathcal{O}(\sqrt{9S+1}\log{\frac{9S+1}{\epsilon}})$, each requiring $\mathcal{O}\big((40S^2+8S)\sqrt{9S+1}\log{\frac{9S+1}{\epsilon}}\big)$ arithmetic operations \cite{nesterov1994interior}. 

The following result establishes the feasibility of $\mathbf{P2}$.

\begin{remark}
    Since the maximum output power of the RF-EH circuit is $g(\varpi_\mathrm{sat})$, $\mathbf{P2}$ is feasible when $\sum_{\forall s \in \mathcal{S}} \frac{E_s^\mathrm{th}}{g(\varpi_\mathrm{sat})} \leq \tau_\mathrm{max}$. Notice that this result is stricter than the feasibility condition of $\mathbf{P1}$, which is feasible when $\tau_\mathrm{max} g(\varpi_\mathrm{sat}) \geq \mathrm{max}_s \ E_s^\mathrm{th}$, as it potentially allows driving the EH circuits at the saturation level during the transmission block.
\end{remark}
\vspace{-3mm}
\subsection{Order-aware time division}
The previous time-beamforming allocation strategy may unintentionally power non-scheduled devices, in which case the approximation \eqref{eq:approxHarvestedEnergy} will no longer hold. Motivated by this, we propose an order-aware time division algorithm in which the power/time allocation corresponding to each device is recomputed after $\mathbf{P2}$ is solved based on the energy they can harvest from others' time-beamforming allocations. Specifically, the PB schedules at each time slot the device whose timeslot-beamforming allocation is the most polluting for other devices. For instance, during the first timeslot, the PB schedules the $s^\mathrm{th}$ device that satisfies
\begin{equation}\label{eq:scheduledDevice}
    s = \underset{s''}{\operatorname{arg\,max}} \sum_{\forall s' \in \mathcal{S}\backslash s''} \tau_{s''} g(\varpi_{s',s''}),
\end{equation}
where $\varpi_{s',s''}$ is the received power at device $s'$ when transmitting the signal vector $\mathbf{w}_{s''}x_{s''}$, and then, we recompute the power allocation for the non-scheduled devices, i.e., $\forall s' \in \mathcal{S}\backslash s$, according to
\begin{align}\label{eq:update1}
    p_{s'} = \frac{N}{\lVert \mathbf{h}_{s'} \rVert_1^2}g^{-1}\left(\frac{\Tilde{E}_{s'}^\mathrm{th}}{\tau_{s'}}\right),
\end{align}
unless $\frac{p_{s'}}{N}\lVert \mathbf{h}_{s'} \rVert_1^2 < \varpi_0$, in which case
\begin{align}
    \Big\{p_{s'} = \frac{\varpi_0 N}{\lVert \mathbf{h}_{s'} \rVert_1^2}, \qquad
    \tau_{s'} = \frac{\Tilde{E}_{s'}^\mathrm{th}}{g(\varpi_0)}\Big\}, \label{eq:update2b}
\end{align}
where $\Tilde{E}_{s'}^\mathrm{th} = E_{s'}^\mathrm{th} - \tau_s g(\varpi_{s',s})$ is the updated energy requirement for the non-scheduled devices. Similarly, we schedule the remaining devices, discarding previously charged ones from the updates. Finally, Algorithm~\ref{alg:tdma} sketches the full procedure.

\begin{algorithm}[t!]
\caption{Order-aware time division}
\begin{algorithmic}[1] \label{alg:tdma}
 \STATE \textbf{Input:} $\{E_s^\mathrm{th}, \mathbf{h}_s\}_{\forall s\in\mathcal{S}}, \tau_\mathrm{max}$ \label{lin1}
 \STATE Solve $\mathbf{P2}$ \label{lin2}
\REPEAT
\label{lin3}
\STATE Schedule the $s^\mathrm{th}$ device satisfying \eqref{eq:scheduledDevice} \label{lin4}
\STATE Update: $\{p_{s'},\tau_{s'},\Tilde{E}_{s'}^\mathrm{th}\} _{\forall s'\in \mathcal{S}\backslash s}$ using \eqref{eq:update1}-\eqref{eq:update2b} \label{lin5}
\STATE Update: $\mathcal{S} \leftarrow \mathcal{S} \backslash \{s\}$
\UNTIL{$|\mathcal{S}| = 1$} \label{lin6}
\STATE Schedule the remaining device in $\mathcal{S}$
\STATE \textbf{Output:} $\{p_{s}, \tau_{s}\}$ for all devices
\end{algorithmic}
\end{algorithm}

\subsection{Charging all-at-once}
In this subsection, we consider the benchmark technique allowing the PB to simultaneously charge all the devices. The PB uses a single analog precoder over the charging time. Hence, we can state the resulting optimization problem as
\begin{subequations}\label{P3}
    \begin{alignat}{3}
    \mathbf{P3:} \quad &\underset{p,\mathbf{w},\tau}{\mathrm{min.}} \quad && p \tau \label{P3a}\\
    &\text{s.t.} \quad && \tau g\left(| \mathbf{h}_s^H \mathbf{w} |^2 \right) \geq E_s^\mathrm{th}, \quad && \forall s \in \cal S, \label{P3b}\\
    & \quad && \tau \leq \tau_\mathrm{max}, \label{P3c}\\
    & \quad && |w_n|^2 = \frac{p}{N}, \ n = 1,\ldots,N. \label{P3d}
    \end{alignat}
\end{subequations}
Since the objective \eqref{P3a} is quasiconcave and the constraints \eqref{P3b} and \eqref{P3d} are non-convex, $\mathbf{P3}$ is a non-convex problem. To circumvent this, we resort to a heuristic method in which the optimal spatial direction is obtained from the following max-min weighted incident RF power (non-convex) problem
\begin{subequations}
    \begin{alignat}{3}
        \mathbf{P4:} \quad &\underset{\mathbf{w}}{\mathrm{max.}} \quad \underset{s}{\mathrm{min}} \ \frac{|\mathbf{h}_s^H \mathbf{w}|^2}{E^\mathrm{th}_s}
    \end{alignat}
\end{subequations}
such that the coefficients $1/E^\mathrm{th}_s$ prioritize the devices with higher energy demands. This guarantees near-optimal results for $\mathbf{P3}$. Then, $\mathbf{P4}$ can be recast into its epigraph form as
\begin{subequations}
    \begin{alignat}{3}
        \quad &\underset{\mathbf{w}}{\mathrm{max.}} \quad && \xi \\
        &\text{s.t.} \quad && |\mathbf{h}^H_s \mathbf{w}|^2 \geq \xi E^\mathrm{th}_s, \quad \forall s \in \mathcal{S}.
    \end{alignat}
\end{subequations}
Next, let us rewrite $|\mathbf{h}^H_s \mathbf{w}|^2 = \mathbf{h}^H_s \mathbf{w} \mathbf{w}^H \mathbf{h}_s = \mathrm{tr}(\mathbf{W} \mathbf{H}_s)$, with $\mathbf{W} = \mathbf{w}\mathbf{w}^H$ and $\mathbf{H}_s = \mathbf{h}_s\mathbf{h}^H_s$ being rank-$1$ matrices, and recast $\mathbf{P4}$ into a semidefinite programming problem
\begin{subequations}\label{P5}
    \begin{alignat}{3}
    \mathbf{P5:} \quad &\underset{\xi,\mathbf{W}}{\mathrm{max.}} \quad && \xi \label{P5a}\\
    &\text{s.t.} \quad && \mathrm{tr}(\mathbf{W} \mathbf{H}_s) \geq \xi E^\mathrm{th}_s, \quad \forall s \in \mathcal{S}, \label{P5b}\\
    & \quad && \mathbf{W} \succeq 0. \label{P5c}
    \end{alignat}
\end{subequations}
Here, we relaxed the rank-$1$ requirement of $\mathbf{W}$ to make the problem convex. Then, we take $\bm{\theta}$, where $\mathbf{w} = \sqrt{\frac{1}{N}}e^{j\bm{\theta}}$, as the phase of the dominant eigenvector of $\mathbf{W}$, which is optimal when $\mathbf{P5}$ yields a rank-$1$ solution.

Then, from the constraint \eqref{P3b}, we derive the optimal power allocation for a given $\tau$ as
\begin{equation}\label{eq:optPowerSDMA}
    p(\tau) = \underset{s}{\mathrm{max}} \ g^{-1}\Big(\frac{E^\mathrm{th}_s}{\tau}\Big) \frac{1}{|\mathbf{h}^H_s \mathbf{w}|^2},
\end{equation}
which is the transmit power required for charging the device with the highest ratio between the needed harvesting power and the input RF power when using the power-normalized precoder $\mathbf{w}$. Finally, we perform an exhaustive search to find the pair $(p,\tau)$ that minimizes the objective function in \eqref{P3a} by evaluating \eqref{eq:optPowerSDMA} over the interval $\tau \in [\frac{\mathrm{max}_s \ E^\mathrm{th}_s}{g(\varpi_\mathrm{sat})}, \tau_\mathrm{max}]$. After computing the optimal power allocation, we update the optimal (analog) precoder as $\mathbf{w}^* = \sqrt{p} \mathbf{w}$.

\section{Practical considerations}\label{sec:practicalConsiderations}
In practice, the resolution of the phase shifters constrains the spatial flexibility of analog beamforming. Hence, our assumption of infinite resolution phase shifters upper bounds the actual performance of both time division and all-at-once schemes when using practical limited-resolution shifters \cite{9721145}. This limitation motivates exploring alternative analog beamforming implementations in which the array pattern is controlled by switches, variable gain amplifiers, or even the induction of a nearby antenna. Notice that a formal comparison of those architectures requires careful consideration of power consumption and costs of the corresponding components. Besides, notice that scheduling groups of devices instead of using our proposed time division strategy becomes more appealing in massive IoT deployments. Therefore, charging one device at a time as the network grows may not suffice the energy demands due to excessive delays. Moreover, notice that our strategy extends to the multi-antenna RF-EH case when the signal is combined in the RF domain. In other cases, such as when the receiver comprises multiple RF-EH circuits, the extension is not straightforward. Finally, notice that our proposed strategy relies on estimating the channel state information at the beginning of the charging block for computing the optimal power and time allocation. However, depending on the coherence time of the channel, the channel state information of the latter scheduled devices may be inaccurate. In this regard, we can rely on a robust analog beamforming formulation \cite{8558720} of our problem or, instead, exploit the channel's statistics \cite{9184149}, which are easier to estimate and vary more slowly. We left the aforementioned concerns as topics for future work.

\section{Numerical results}\label{sec:numericalResults}
Here, we evaluate the PB energy consumption under the proposed time division schemes, referred to as order-agnostic and order-aware for short, against the all-at-once strategy.\footnote{For the all-at-once strategy, we employ a linear interpolant method to fit $g(\cdot)$ (and its inverse) to the Powercast RF-EH circuit data, allowing us to evaluate \eqref{eq:optPowerSDMA} and adjust the beamforming power of the proposed time division strategies, to ensure a fair comparison.}
\begin{figure}[t!]
    \centering
    \includegraphics[width = .5\columnwidth]{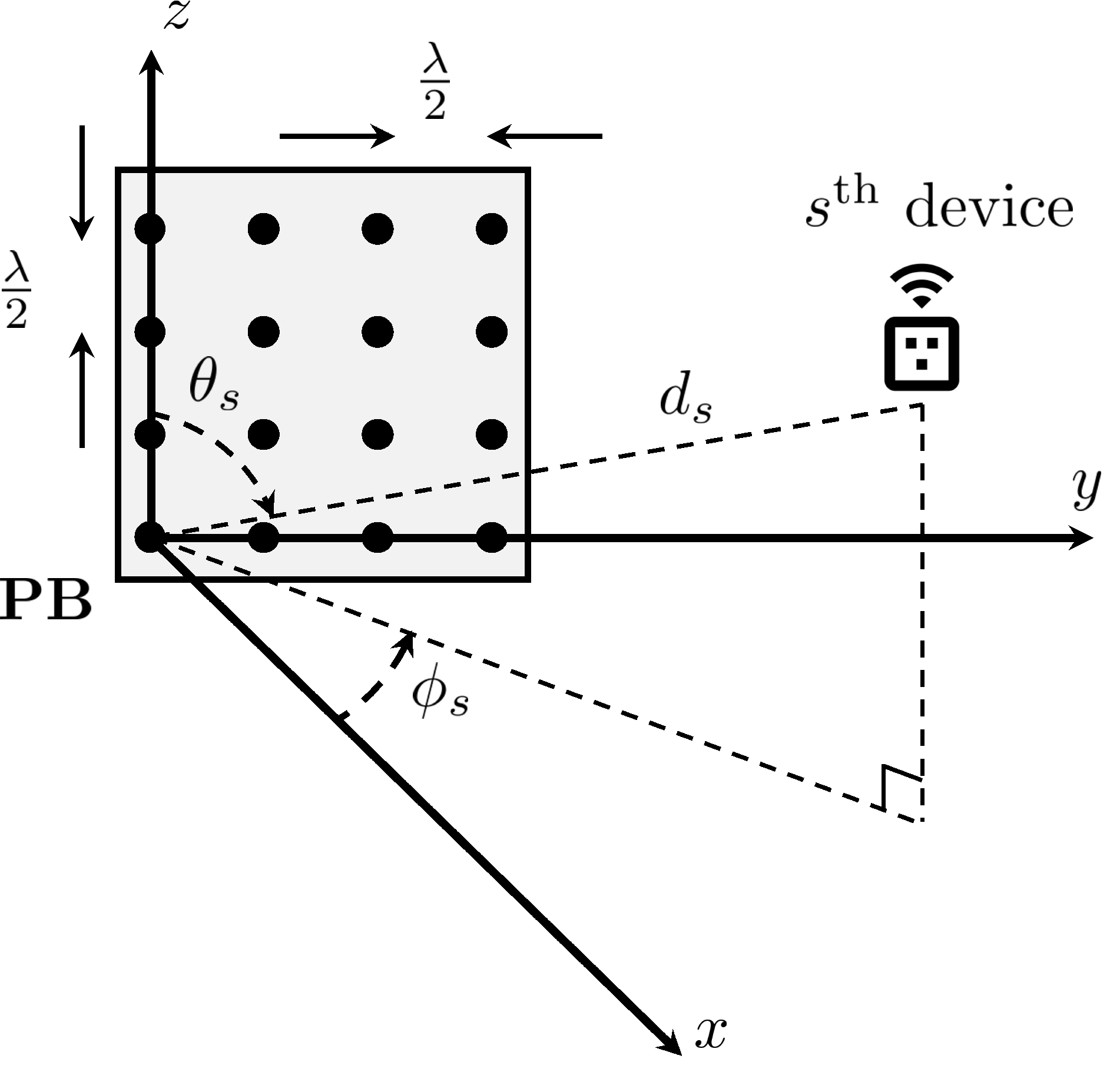}
    \vspace{-.8em}
    \caption{Geometry of the simulation setup. The antenna elements are separated by half the wavelength of the transmitted signal in both $z$ and $y$ directions.}
    \vspace{-3mm}
    \label{fig:simulationSetup}
\end{figure}

As Fig.~\ref{fig:simulationSetup} illustrates, the PB, deployed at the origin, is equipped with a uniform rectangular array with the same number of antennas in both horizontal and vertical dimensions, i.e., $\sqrt{N}$. Meanwhile, the path-loss for the $s^\mathrm{th}$ user located at a distance $d_s$ from the PB is modeled as $\beta_s = 10^{-1.6} d_s^{-2.7}$ \cite{goldsmith2005wireless}, while the channel is subject to Rician fading, thus
\begin{equation}
    \mathbf{h}_s = \sqrt{\frac{\kappa \beta_s}{1 + \kappa}} \mathbf{h}_s^\mathrm{los} + \sqrt{\frac{\beta_s}{2(1+\kappa)}} \mathbf{h}_s^\mathrm{nlos},
\end{equation}
where $\mathbf{h}_s^\mathrm{los}$ and $\mathbf{h}_s^\mathrm{nlos}$ are the line-of-sight (LoS) and the NLoS channel components. Moreover, the LoS factor $\kappa$ is set to $10$ and we model the LoS component of the channel as $\mathbf{h}_s^\mathrm{los} = \mathbf{h}_s^h \otimes \mathbf{h}_s^v$, where $\mathbf{h}_s^h \!=\! [1, e^{j\pi \sin{\theta_s} \sin{\phi_s}}, \ldots, e^{j\pi (\sqrt{N} - 1) \sin{\theta_s} \sin{\phi_s}}
\!]^T\!$, $\mathbf{h}_s^v \!=\! \![\! 1, e^{j\pi \cos{\theta_s}}, \ldots, e^{j\pi (\sqrt{N} - 1) \cos{\theta_s}} \!]^T\!$ are the horizontal and vertical LoS components, respectively, for a given elevation angle $\theta_s$ and azimuthal angle $\phi_s$ \cite{balanis2015antenna}. Finally, unless stated otherwise, we consider a setup with three IoT devices respectively located at $\mathbf{v}_1 = [0, \ -5, \ 0]^T$, $\mathbf{v}_2 = [5, \ 0, \ -5]^T$, and $\mathbf{v}_3 = [10, \ 10, \ 5]^T$ (all distances in meters), given Cartesian coordinates, and $E^\mathrm{th}_s = 4~\text{mJ}, \forall s \in \mathcal{S}$. 

\begin{figure}[t!]
    \centering
    \includegraphics[width = 0.9\columnwidth]{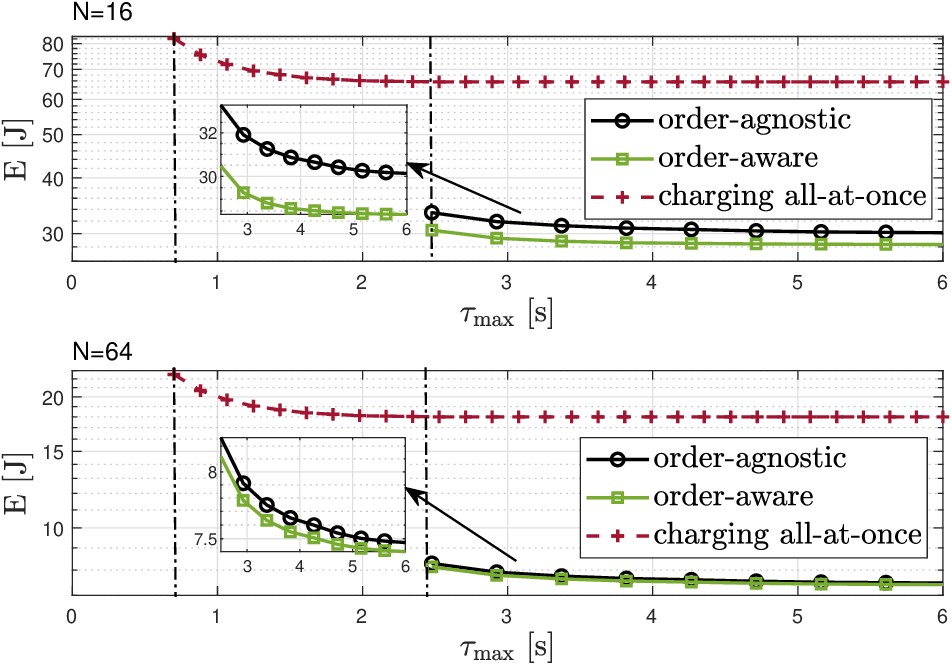}
    \vspace{-1em}
    \caption{PB's energy consumption vs $\tau_\mathrm{max}$ for the time division and all-at-once charging strategies and $N \in \{16, 64\}$. The vertical dash-dotted lines delimit the feasible regions for both approaches.}
    \vspace{-3mm}
    \label{fig:energy_vs_time}
\end{figure}

Fig.~\ref{fig:energy_vs_time} shows the PB's energy consumption (hereinafter denoted as $E$) vs the maximum charging time $\tau_\mathrm{max}$ for all strategies. As expected, the all-at-once strategy provides the shortest charging time, which is approximately $\frac{\mathrm{max}_s E^\mathrm{th}_s}{g(\varpi_\mathrm{sat})}$, since all devices are charged simultaneously. However, once $\mathbf{P2}$ becomes feasible the order-agnostic outperforms the all-at-once strategy since scheduling one device at a time maximizes its achievable conversion efficiency. Further improvements are obtained by exploiting our order-aware scheduling (at the expense of a slight increase in computational complexity), specially when $N = 16$, showing that non-scheduled devices can also harvest energy. However, as the number of antennas increases, the energy beam narrows which reduces the amount of interference at non-scheduled devices. Notice that the energy consumption curves flatten for large values of $\tau_\mathrm{max}$. In such a case, all the RF-EH circuits operate at their maximum efficiency input power $\varpi_0$ when using the time division strategies. Meanwhile, for the all-at-once strategy, only the worst-performing device according to \eqref{eq:optPowerSDMA} is guaranteed to operate nearly at maximum conversion efficiency.

\begin{figure}
    \centering
    \includegraphics[width=0.9\columnwidth]{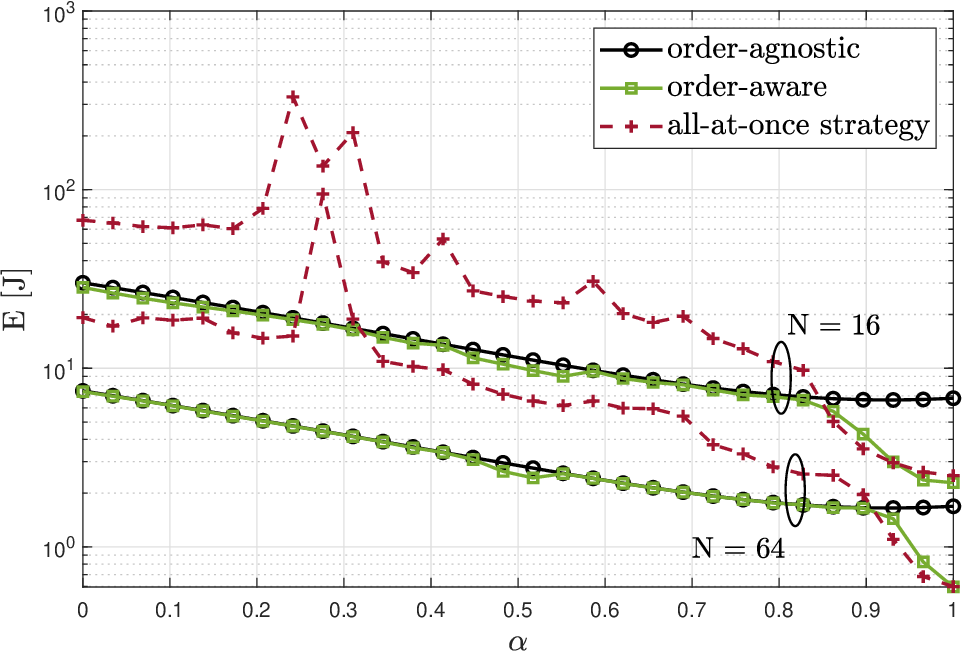}
    \vspace{-1em}
    \caption{PB's energy consumption vs $\alpha$ for $N = \{16, 64\}$, and $\tau_\mathrm{max} = 8~\text{s}$.}
    \vspace{-3mm}
    \label{fig:energy_vs_deployments}
\end{figure}

Fig.~\ref{fig:energy_vs_deployments} shows how the network deployment, captured by the parameter $\alpha$, can change the performance gap among all strategies.
Mathematically, the devices' coordinates are updated for each value of $\alpha\in [0,1]$ as $(1-\alpha)\mathbf{v}_s+\alpha [2.5,\ 2.5,\ 2.5]^T$ such that all the devices approach the common spatial point $[2.5,\ 2.5,\ 2.5]^T$ as $\alpha$ increases. Notice that, increasing the number of antennas reduces the performance gap between the time division strategies as a result of a narrower energy beam. Moreover, when all devices are deployed near each other, around $\alpha = 0.9$, the order-aware scheduling and the all-at-once strategies significantly outperform the order-agnostic scheduling approach. The poorer performance of the order-agnostic scheduling is due to the non-negligible harvested energy at the non-scheduled devices, which is not accurate when all devices are deployed close to each other, separated explicitly by sub-wavelength distances, and a single beam efficiently charges all devices simultaneously. Additionally, observe that under such conditions even the order-aware can become sub-optimal with respect to the all-at-once strategy, specially when $N=64$ for which harvesting from non-intentional charging becomes less feasible.

\section{Conclusions}\label{sec:conclusions}
We proposed an energy-efficient time division RF-WET strategy to charge an IoT network using analog beamforming. To this end, we demonstrated that for the point-to-point RF-WET scenario fitting the transfer function of the RF-EH circuit in the interval between maximum conversion efficiency and saturation suffices for computing the optimal power/time allocation. Results evinced the outstanding performance of the proposed strategy, which improves as the number of transmit antennas increases. Nevertheless, we acknowledge the limitations of this scheme when the network has urgent energy demands, or the number of devices grows massive. Moreover, our findings show that all-at-once strategy is appealing when the devices are tightly spatially clustered. As a result, we leave the problem of developing a scheduling algorithm combining both strategies' advantages as future work.

\textit{Reproducible research}: The simulation results can
be reproduced using the Matlab code available at:
\url{https://github.com/Osmel-dev/EE-analog-beamforming-WET}
\vspace{-.5em}
\bibliographystyle{IEEEtran}
\bibliography{IEEEabrv,references}
\end{document}